\documentclass[conference,10pt]{IEEEtran}
\IEEEoverridecommandlockouts
\PassOptionsToPackage{hyphens}{url}
\usepackage{hyperref,verbatim}
\usepackage{cite}
\usepackage{mwe}
\usepackage{subcaption}
\usepackage{multirow}
\usepackage{amsmath,amssymb,amsfonts}
\usepackage{algorithmic}
\usepackage{graphicx}
\usepackage{textcomp}
\usepackage{xcolor}
\def\BibTeX{{\rm B\kern-.05em{\sc i\kern-.025em b}\kern-.08em
    T\kern-.1667em\lower.7ex\hbox{E}\kern-.125emX}}

\renewcommand{\baselinestretch}{0.965}
    
\begin{document}

\title{A 9 Transistor SRAM Featuring Array-level XOR Parallelism with Secure Data Toggling Operation}

\author{\IEEEauthorblockN{Zihan Yin$^{1,2}$, Annewsha Datta$^{1,2}$, Shwetha Vijayakumar$^{1}$, Ajey Jacob$^{1,2}$, Akhilesh Jaiswal$^{1,2}$}
\it{$^1$University of Southern California, Los Angeles, USA \qquad $^2$Information Sciences Institute, Marina Del Rey, USA} \\
 	Correspondence: \it{zihanyin@usc.edu} \\[-1.0ex]}

\maketitle
\begin{abstract}
Security and energy-efficiency are critical for computing applications in general and for edge applications in particular. Digital in-Memory Computing (IMC) in SRAM cells have widely been studied to accelerate inference tasks to maximize both throughput and energy efficiency for intelligent computing at the edge. XOR operations have been of particular interest due to their wide applicability in numerous applications that include binary neural networks and encryption. However, existing IMC circuits for XOR acceleration are limited to two rows in a memory array and extending the XOR parallelism to multiple rows in an SRAM array has remained elusive. Further, SRAM is prone to both data imprinting and data remanence security issues, which poses limitations on security . 
Based on commerical Globalfoundries 22nm mode, we are proposing a novel 9T SRAM cell such that multiple rows of data (entire array) can be XORed in a massively parallel single cycle fashion. The new cell also supports data-toggling within the SRAM cell efficiently to circumvent imprinting attacks and erase the SRAM value in case of remanence attack.


\end{abstract}
\vspace{2mm}

\section{Introduction}

\IEEEPARstart{A}{rtificial} Intelligence of Things (AIoT) combines AI with IoT infrastructure for ubiquitous data analytics\cite{zhao2022event}, requiring energy-efficient, fast solutions due to resource scarcity and limited energy in IoT platforms \cite{liu2020ns}. Binarized neural networks (BNNs) offer reduced model size, compute requirements, and energy consumption, yielding hardware savings and computation precision \cite{zhang202155nm,8778160}. They rely heavily on XOR operations, which need hardware acceleration for swift, energy-efficient computations within resource limits.

On the other hand, SRAM, a high-speed memory type often used for CPU cache, is crucial in digital systems, especially AIoT applications. While traditionally used for storage, executing XOR operations, vital in BNNs of AIoT systems, within SRAM cells can overcome the memory-wall bottleneck \cite{9382915}. This integration can yield a more efficient system for BNN's computational demands and pave the way for AIoT-focused CPUs, with improved performance and energy efficiency. Hence, the deep interplay between AIoT and CPU components like SRAM highlights the need for enabling massively parallel XOR operations within the SRAM arrays.

Further, the globalization of semiconductor design and fabrication increases the vulnerability of integrated circuits (ICs) to threats, as for SRAM, it often stores sensitive data, which if leaked, could compromise system security. Emerging memory attacks, like the cold boot\cite{yitbarek2017cold,halderman2009lest} and Negative Bias Temperature Instability (NBTI) aging attacks\cite{paul2005impact,furusawa2014high}, exploit hardware vulnerabilities and data remanence properties of SRAM, allowing unauthorized data access even on encrypted systems. The NBTI aging attack, a MOSFET reliability issue, intensifies with transistor miniaturization and can lead to data imprinting attacks\cite{ho2014dynamic,ho2019secure}. These vulnerabilities can risk security in numerous applications, including machine learning, by exposing trained model parameters through unsecured SRAM arrays. Therefore, developing robust countermeasures to bolster SRAM security and mitigate its inherent vulnerabilities is essential, with a focus on data remanence and data imprinting hardware vulnerabilities in SRAM arrays.

 \textit{Prior-Works:} Due to the inherent computational benefits of BNN, optimal hardware architecture based on In-Memory Computing (IMC) with analog-domain and digital computation are actively sought to minimize the excessive data movement energy to/from on-chip memories\cite{7875410,8267253,9035482}. However, prior works \cite{liu2018parallelizing}\cite{8401845} on acceleration of digital in-memory XOR operations have been limited to only two rows in a memory array as compared to our proposed  design which can XOR more than 2 rows (entire memory array), simultaneously. Further, for solving the problem of SRAM data attacks\cite{9256595}, prior-art designs use very large SRAM cells consisting of 22\cite{ho2019secure} and 15 transistors\cite{ho2016area} per cell which costs large area and power overheads.

In order to address the aforementioned challenges, we introduce an innovative circuit configuration comprised of a 9T SRAM cell. This design facilitates the exclusive OR (XOR) operation across multiple data rows (entire array) in a highly parallelized manner, thereby enhancing computational efficiency. Additionally, the proposed design allows for the efficient toggling of data within the SRAM cell, a feature that not only bolsters the security of the cell, thus reducing vulnerability to unauthorized data access but also extends its operational lifespan. This dual improvement in security and longevity underscores the utility and potential of our innovative 9T SRAM cell design. The proposed SRAM cell can also be used for high-speed data erasing in case of a data remanence threat. \textit{Thus, our proposed SRAM cell can enable massively parallel array-level in-memory XOR operation as well as immunity to SRAM imprinting and remanence attacks.}

\begin{figure*}[ht]
\centering
\begin{subfigure}{0.3\textwidth}
\centering
\includegraphics[width = \textwidth]{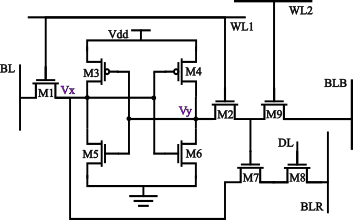}
\caption{9T SRAM Circuit Design}
\label{fig:9T}
\end{subfigure}
\hfill
\begin{subfigure}{0.3\textwidth}
\centering
\includegraphics[width = \textwidth]{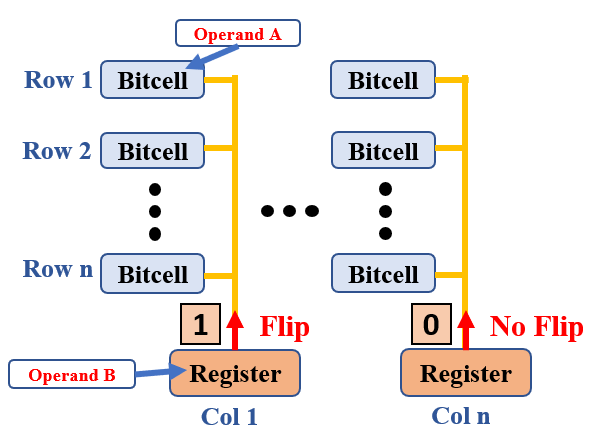}
\caption{Block Diagram of the SRAM Array in XOR Mode}
\label{fig:XOR}
\end{subfigure}
\hfill
\begin{subfigure}{0.3\textwidth}
\centering
\includegraphics[width = \textwidth]{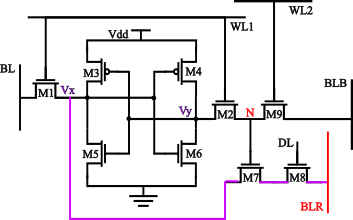}
\caption{Circuit diagram of the bitcell in XOR Mode}
\label{fig:XOR}
\end{subfigure}
\caption{Proposed 9T SRAM block and circuit diagrams}
\label{XORf}
\end{figure*}

\section{Circuit design}
\label{II}
The proposed 9T SRAM design is depicted in Fig. \ref{fig:9T}, which consists of the 6T SRAM element (transistors M1-M6) and additional transistors M7, M8 and M9 for supporting XOR operations and flipping the value stored in the bit-cell, as required for multi-row XOR and massively parallel data-toggling operations.

The proposed 9T SRAM can work in the normal SRAM read and write mode (similar to conventional 6T SRAM) as well as in massively parallel array-level XOR operation mode.
\subsection{Normal Mode}
When functioning as in normal SRAM mode: the DL is OFF and WL1 and WL2 are ON and used as conventional Word-line for differential read and write as shown in Fig. 
\ref{XORf}(a).

\begin{enumerate}
    \item When performing the write operation, the BL and BLB will have opposite values `1' and `0' (depending on the data to be written) while both WL1 and WL2 would be kept ON to allow the read/write current to pass through transistors M2 and M9 on the right and M1 on the left to charge/discharge the voltage of nodes $Vx$ and $Vy$ to the desired value VDD or GND. 
    \item  When performing the read operation both WL1 and WL2 are ON and the voltage is read through BL and BLB through a column sense-amplifier.
\end{enumerate}
Note that the new bit-cell circuit with the added 3 transistors is asymmetric compared with the conventional 6T SRAM circuit. As we will show later, the asymmetry in an appropriately designed proposed 9T SRAM cells only leads to minimal change in static read and write noise margins compared to a symmetric 6T SRAM design.

\subsection{XOR Mode}
\label{subsecxor}
In this subsection, we would now describe a method of implementing the XOR operation within the 9T cell shown in Fig. \ref{XORf}(a). Then we would also describe how the parallelism of XOR operation can be modified to include multiple rows at the same time resulting in massively parallel array-level in-memory operation.

\begin{table}[!t]
\centering
\caption{Truth Table of the XOR}
\begin{tabular}{|ll|l|l|} \hline
A & B & OUT &   \\ \hline
\hline
0 & 0 & 0 & OUT = A \\
0 & 1 & 1 & OUT = not A  \\
1 & 0 & 1 & OUT = A \\
1 & 1 & 0 & OUT = not A \\ \hline
\end{tabular}
\label{t1}
\end{table}

For the proposed XOR operation, let us first assume that the memory array is arranged in a 2-D pattern consisting of rows and columns as shown in Fig. \ref{XORf}(b). We consider the operand A for the XOR operation is stored in the bitcell. Further, let us consider it is stored in the first top left bitcell of Row 1. The operand B that has to be XORed with operand A is stored in the bottom left registers in \ref{XORf}(b) that is outside of the memory array. Referring to Table \ref{t1}, our designed circuit ensures operand A and B are XORed and the result is overwritten in the bitcell storing operand A. And for each bitcell (Row 1 to Row n) in column 1, the XOR operation between their respective operand A values stored within the respective bitcells and the operand B stored in the register outside the memory array in column 1 executes, simultaneously, which enables massive parallel multi-row XOR operation. 

There are two steps involved in performing XOR operation within the proposed bitcell shown in Fig. \ref{fig:XOR}. In step 1, we implement a global reset (\textit{i.e.}, $Vx="0"$, $Vy="1"$) through BLR while keeping the original value of the bitcell on the dynamic node N. In step 2, for cases in which operand B in Table \ref{t1} is `1', implying that bit-cell value has to be inverted to accomplish XOR operation, the circuit leverages line BLR to flip the value inside the bitcell to achieve the desired XOR operation. 
\begin{figure*}[!th]
\centering
\includegraphics[width = \linewidth ]{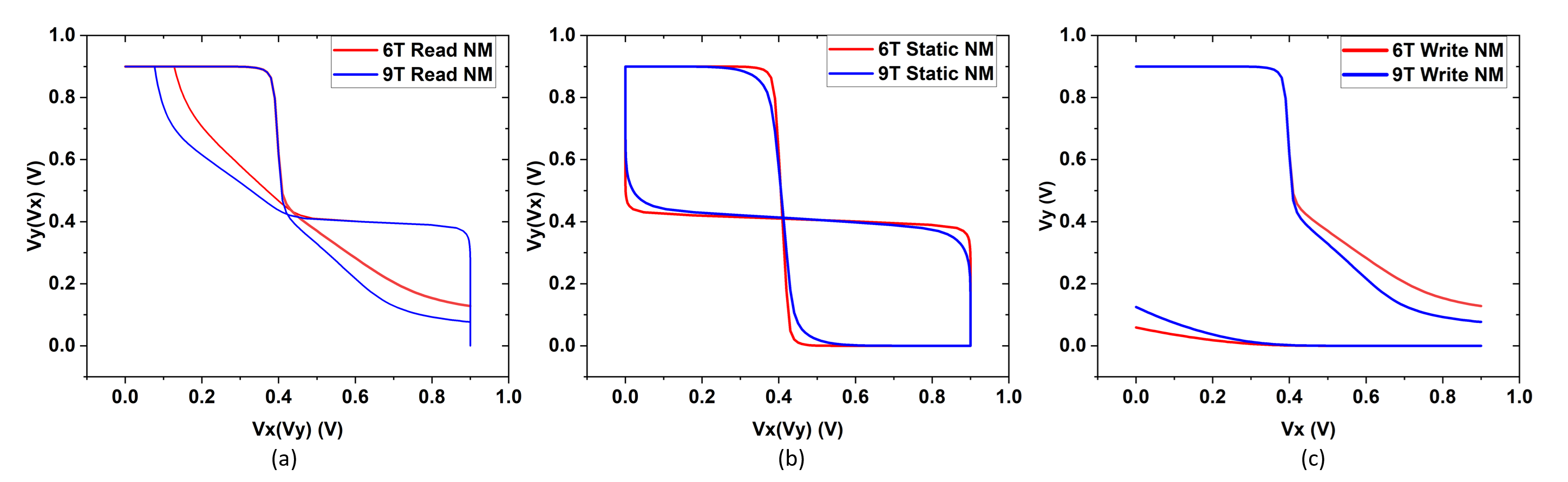}
\caption{6T SRAM and the proposed 9T Secure SRAM's static, write and read noise margin comparisons.}
\label{fig:NM}
\vspace{-3mm}
\end{figure*}
\begin{figure*}[th]
\centering
\begin{subfigure}{0.45\linewidth}
\centering
\includegraphics[width = \textwidth]{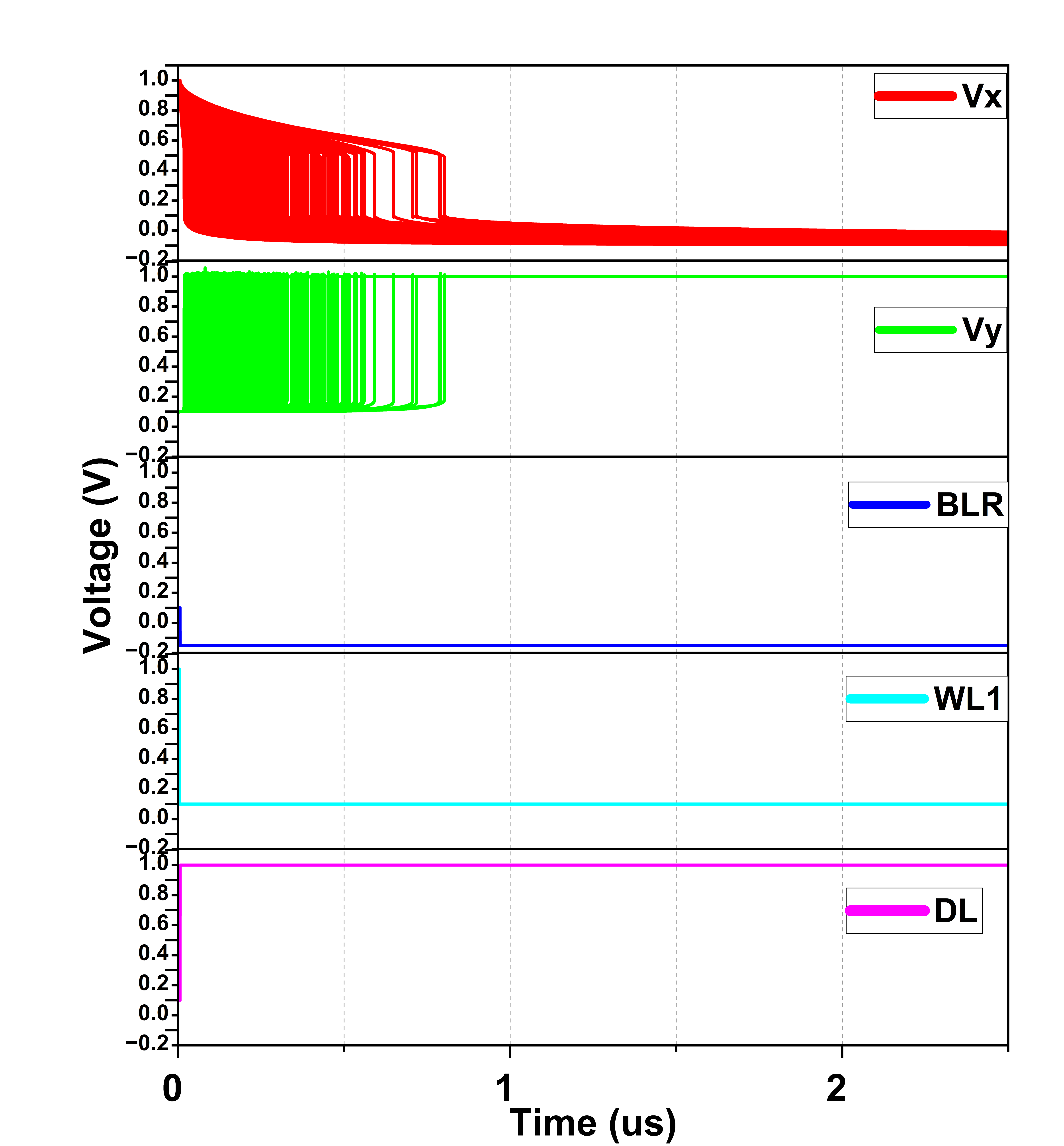}
\caption{}
\label{fig:sim1}
\end{subfigure}
\begin{subfigure}{0.45\linewidth}
\centering
\includegraphics[width = \textwidth]{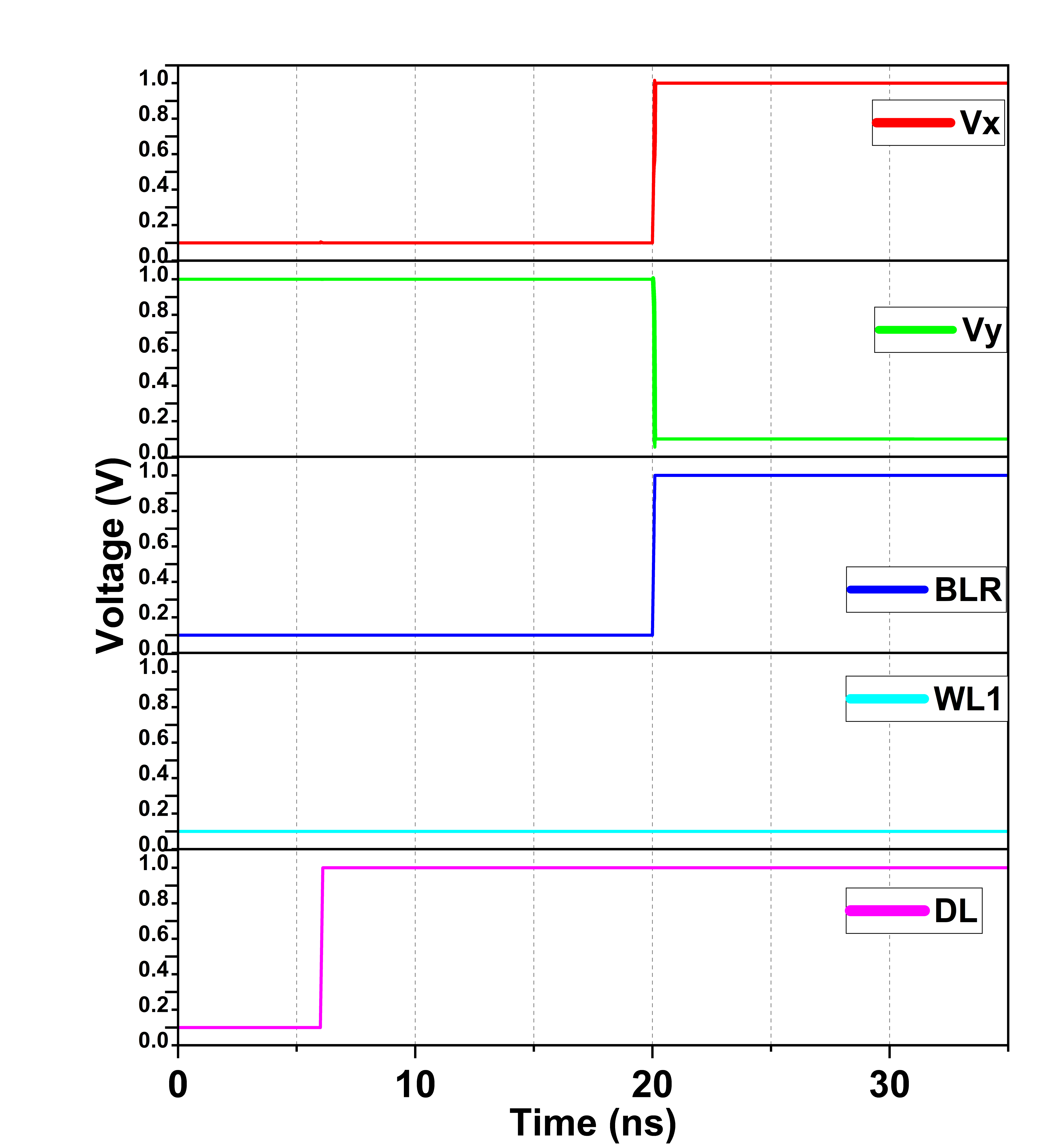}
\caption{}
\label{fig:sim2}
\end{subfigure}
\caption{(a) is the Monte Carlo simulation result of the proposed 9T SRAM circuit in XOR Mode step 1, testing the case of operand A is '1' and operand B is '1' and (b) is the Monte Carlo simulation result of the proposed 9T SRAM circuit in XOR Mode Step 2, testing case of operand A is '0' and operand B is '1'.}
\label{sim}
\end{figure*}
\begin{table*}
\caption{XOR Mode Two-Step Circuit Node Value}
\centering
\begin{tabular}{|c|c|c|c|c|c|c|c|c|c|c|c|c|c|} 
\hline
\multirow{2}{*}{\textbf{Case}} & \multirow{2}{*}{\textbf{Operand A}} & \multirow{2}{*}{\textbf{Operand B}} & \multirow{2}{*}{\textbf{Node N}} & \multirow{2}{*}{\textbf{M7}} & \multicolumn{4}{c|}{\textbf{Step 1}} & \multicolumn{4}{c|}{\textbf{Step 2}} & \multirow{2}{*}{\textbf{Bitcell Result}} \\ 
\cline{6-13}
 &  &  &  &  & \multicolumn{1}{l|}{\textbf{\pmb{$V_x$}}} & \multicolumn{1}{l|}{\pmb{$V_y$}} & \textbf{DL} & \textbf{BLR} & \pmb{$V_x$} & \multicolumn{1}{l|}{\pmb{$V_y$}} & \multicolumn{1}{l|}{\textbf{DL}} & \textbf{BLR} &  \\ 
\hline
\hline
Case 1 & 0 & 0 & 1 & ON & 0-0 & 1-1 & 0 & 0
  & 0-0 & 1-1 & 0 & 0 & 0 \\ 
\hline
Case 2 & 0 & 1 & 1 & ON & 0-0 & 1-1 & 1 & Negative
  voltage & 0-1 & 1-0 & 1 & 1 & 1 \\ 
\hline
Case 3 & 1 & 0 & 0 & OFF & 1-1 & 0-0 & 0 & 0
   & 1-1 & 0-0 & 0 & 0 & 1 \\ 
\hline
Case 4 & 1 & 1 & 0 & OFF & 1-0 & 0-1 & 1 & Negative
  voltage & 0-0 & 1-1 & 1 & 1 & 0 \\
\hline
\end{tabular}
\label{t2}
\end{table*}

\begin{enumerate}
    \item Step 1: $Conditional \ Reset$: During this step, let us first assume that the node $V_x$ is at logic '1' and $V_y$ is at logic '0' implying that the bitcell is storing a 1 representing the value of Operand A,  while operand B is '1' and is stored in the register outside the array, WL1 is pulled up to VDD so both transistor M1 and M2 are switched ON. WL2 is kept at GND. Since we are not reading or writing the data in the SRAM cell, both BL and BLB should be pre-charged. When M2 is switched ON, while keeping M9 OFF, voltage on node N settles down to the same voltage as $V_y$. As a result, the gate voltage of M7 is discharged to logic '0'. Then we pull down line WL1, so both M1 and M2 are OFF. Now we apply a negative voltage to BLR and DL would have the same value of Operand B which in this case is assumed to be '1', DL is pulled up to VDD and the transistor M8 is switched ON. As M8 is open and line BLR is driven at a negative voltage, node $V_x$ can be discharged from the part consisting of $V_x$, M7, M8 (as shown in the pink line in Fig. \ref{XORf}(c)) to line BLR. This causes the node $V_x$ to experience significant discharge, subsequently the cross-coupled inverter ensures the voltage of the node $V_x$ settles down to GND. (Note that transistors M7, M8 are made strong to ensure the bitcell flips through M7 and M8 and is reset to '0'.) $V_x$ is then at logic '0' and $V_y$ at logic '1' while node N stores the original value ('0') of the operand A as a dynamic voltage. \\
    In the case when operand A is '1' and operand B is '0'. Since DL is pulled to the same value as B, it stays at GND. Thus, node N is at GND as well as DL. The negative voltage of BLR could not pass through the series of transistors M7 and M8. As the gate of both the transistors is at GND. Consequently, node $V_x$ stays at the original logic value of '1' after the reset step.\\
    In the other two cases wherein initially $V_x$ is at logic '0', node N would store '1' since node $V_y$ is originally at logic '1'. The voltage at node $V_x$ would stay at value '0' irrespective of the voltage at DL and the negative voltage at BLR.
    
    \item Step 2: $Conditional\ Flip$: For the step 2, as shown in Table \ref{t1} the bit-cell data needs to flip based on the operand B (i.e. if operand B is `1'). We will describe the four cases of the XOR truth Table separately. Let us first consider the case in which operand A is '0' and operand B is '1' so the XORed result should be '1'. After the reset in step 1, the bitcell's node $V_x$ is at logic '0',  so in accordance to the truth table of the XOR logic, $V_x$ has to be flipped to '1'. Before step 2, dynamic node N stores the logic '1'. During step 2, WL1 is kept at at GND and M7 stays ON due to high voltage at node N. Line DL is kept at the same voltage level as operand B which in this case, is pulled up to VDD as operand B is at logic '1'. Also, line BLR is driven by the same voltage of DL which in this case is at VDD. As the voltage at node N and DL are both high, transistors M7 and M8 are switched ON. So node $V_x$ is pulled up to logic '1' through BLR through transistors M7, M8 as shown by the pink line in Fig. \ref{XORf}(c), the bit-cell flips making $V_x = '1'$. $V_x$ stores the XORed result of operands A and B after step 2 finishes, corresponding to Case 2 in Table \ref{t2}.\\
    In the two cases of operand B is at '0', the value at $V_x$ and $V_y$ does not change since during step 2 both DL and BLR stay at GND so the transistor M8 is OFF. The voltage at node $V_x$ cannot be passed through transistor M8 to GND at BLR.\\
    In the last case of operand A at '1' and operand B at '1'. After step 1 the node $V_x$ is reset to logic '0' while node N stores the temporary '0' as node $V_y$. So during step 2, as both BLR and DL are at VDD as operand B is at logic '1'. Transistor M7 is switched OFF and node $V_x$ would keep the reset value '0' after the operation. In summary the bitcell value which is stored at node $V_x$ is modified after Step 2 to store the correct XOR value for all 4 cases of XOR operation of Operand A and B. Table \ref{t2} shows the circuit nodes values during all 4 cases of XOR operation and operand A and B in both steps.
    
\end{enumerate}

\subsection{Massive Parallel XOR Operation}

For massive parallel XOR operation in the proposed circuit block as shown in Fig. \ref{XORf}(b). Take the operation on Column 1 as an example. The data for operand A are stored within the bitcells while operand B is applied from outside the array to the register that is at the bottom of Column 1. During XOR mode, in accordance with Table \ref{t2}, when the operand B value is set to logic '1' the bitcell result would be flipped compared to the original bitcell value which is set as operand A. To achieve this, we would first activate the WL1 shown in Fig. \ref{fig:9T} in those rows (could be more than 2 rows) of the array that needs to be XORed. And then perform the two steps as mentioned in section \ref{subsecxor}. Note that the yellow line in Fig. \ref{XORf}(b) is the line BLR in  both Fig. \ref{XORf}(a) and (c). As shown in the figure, all bitcells within one column connect to one single BLR line. Similarly, all WL lines in Fig. \ref{XORf}(a), within one row connects to the same WL line. While we are executing the flipping operation on bitcells in a particular row, other bitcells  in other WL activated rows will also flip their stored value for those columns whose operand B's value is logic '1'. And all of these flipping operation is performed simultaneously in one cycle. Thus, with the proposed bitcell circuit design we can execute the XOR operation with one operand (operand B in Fig. \ref{XORf}(b)) with more than one rows in the memory array at the same time. In fact, we can execute the XOR operations of the entire memory array simultaneously within one cycle with respect to a given vector operand B. For example, operand B could be binary activations for a BNN that needs to be XORed with multiple weight filters stored in the memory array as multiple operand A. Similarly, operand B could be data to be encrypted while A being the the encryption key to be XORed with the data.

\subsection{Data Toggling Mode}
Following up on the last section, while operating massive parallel XOR operations. If operand B for all columns is in logic '1', then the entire memory array will flip in one cycle. This result enables the idea of data toggling mode. The data toggling is achieved by performing XOR with an arbitrary 1. As shown in Fig. \ref{XORf}(b), by providing logic value '1' to all the registers at the bottom of all the columns (i.e. operand B), then as shown in Table \ref{t2}'s case 2 and 4, the bitcell result will flip after the two-step operation described in section \ref{subsecxor}. This would result in massively parallel data toggle operation, that can embed immunity against data imprinting attacks on SRAM, by enabling low-overhead frequent data toggle operation to avoid asymmetric in NBTI aging for the SRAM bitcell.

\subsection{Erase Mode}
\label{EM}
Along with the data-toggling function, step 1 of the XOR mode in \ref{subsecxor} can also be used for erasing data. This is because at step 1, when operand B is a logic '1', despite the value of operand A, at the end of step 1, node $V_x$'s value is at logic '0' and $V_y$ at logic '1' which acts as the reset of the bitcell. Therefore, providing a logic '1' to operand B in every column of the SRAM array could be regarded as a massive reset signal that enables writing a logic '0' to the $V_x$ of the SRAM cell.

\section{Simulation Results, Analysis and Discussion}
\label{III}

First, Fig. \ref{fig:NM} represents the read, static and write noise margin comparisons between the proposed 9T secure SRAM and the base 6T SRAM cell. It is shown that with the additional three extra transistors added to the base 6T cell, the noise margin is not been affected on a large scale. Especially as in Fig. \ref{fig:NM}(b), the static noise margin is very close to the base 6T's noise margin line. The ability to retain its value inside the cell in the presence of external noise is at the same level as the baseline 6T SRAM cell. As for write operation, from Fig. \ref{fig:NM}(c), additional transistors in the bitcell do not have significant affect on its write-ability which is desirable.

 As shown in section \ref{II}, the proposed 9T secure SRAM circuit supports the XOR mode, enabling massive parallel XOR operation. Fig. \ref{fig:sim1} and Fig. \ref{fig:sim2} show the Monte Carlo simulation result of the two separate steps in the XOR mode. Fig. \ref{fig:sim1} represents the first step of operation for the XOR mode. As shown in Table \ref{t2}, in step 1, only for the case of operand A and operand B both at logic '1', node $V_x$ requires to flip its original value '1' to '0'. In the other three cases, the node $V_x$ retains its original value. Fig. \ref{fig:sim1} presents simulation results for this specific case. Voltages waveforms for other cases not shown in the plot retain their their VDD or GND value in our simulation results. With 1000 points' Monte Carlo simulation result, it could be seen that when the line BLR is given a negative voltage, after a certain period (1us in the case of the -0.17V negative voltage on line BLR), node $V_x$ is successfully flipped to logic '0' from the original value '1'. Note, this time varies with different negative voltages on line BLR. Fig. \ref{fig:sim2} is the Monte Carlo simulation results of the XOR mode's step 2: the flipping stage. According to Table \ref{t2}, in step 2, only in the second case in which the operand A is '0' and operand B is '1', the result node $V_X$ is been flipped from logic '0' to '1'. Fig. \ref{fig:sim2} represents this case. In this step, line WL1 is kept at GND '0' and node $V_x$ and node $V_y$'s values are flipped after line BLR's voltage is pulled up to VDD.

For the functionality of data-toggling: with additional components in the periphery circuit that could control the input of operand B of each column of the secure SRAM memory, the stored data could be flipped simulatneosuly in the same cycle. When reading the data-toggled data, one extra step of data reversing is needed before further transmission to the CPU. The data toggling operation would be very helpful to reduce the risk of on-chip data theft by physical imprinting attack on SRAM cells. 

Additionally, for the erasing mode mentioned in section \ref{EM}, note that this functionality is achieved by only executing step 1: $conditional \ reset$ of the XOR mode operation. 


\section{Conclusions}
\label{IV}
In this work, we have presented a novel 9T SRAM bitcell design with the ability to perform massive in-memory digital XOR operations within the memory array. Compared to existing SRAM based digital in-memory computing works that are limited to digital XOR of two rows in the memory array our proposal can lead to XOR of more than 2 rows (entire memory array) in a single cycle. Further, it supports data-toggling and data-erasing functionality which can embed immunity to data imprinting and remanence attack on SRAM array. Our proposal is supported by simulation results based on commercial 22nm Globalfoundries technology node. The proposed SRAM bitcell can thus enable \textit{massively parallel `secure' in-memory computing} operations required for wide range on applications.

{\small
\bibliographystyle{ieeetr}
\bibliography{egbib}
}

\end{document}